\documentclass[review]{elsarticle}

\usepackage{lineno,hyperref,graphicx,float}
\usepackage{caption,subfig}
\usepackage{amsmath,amssymb}
\usepackage{makecell}
\usepackage{multirow}
\modulolinenumbers[5]

\journal{}

\begin{document}
	
\captionsetup[figure]{labelformat={default},labelsep=period,name={Fig.}}
%\captionsetup[figure]{labelfont={bf},labelformat={default},labelsep=period,name={Fig.}}
%\newcommand{\reffig}[1]{Figure \ref{#1}}

\begin{frontmatter}

%\title{Robust reversible data hiding in encrypted images using polynomial arithmetic}%
\title{A novel reversible data hiding in encrypted images based on polynomial arithmetic}

\author[1]{Lin Chen}
\author[1]{Jianzhu Lu\corref{cor1}}
\ead{tljz@jnu.edu.cn}
\author[1]{Junguang Huang} 
\author[1]{Huiping Hu}
\author[1]{Jiali Peng}

\cortext[cor1]{Corresponding author}

\address[1]{Department of Computer Science, Jinan University, Guangzhou, 510630, China}

\begin{abstract}
Reversible data hiding in encrypted images is an effective technology for data hiding and protecting image privacy. Although there are many high-capacity methods have been presented in recent year, most of them need a pre-processing phase to reserve room in the original image before encryption. It may be unpractical, because the image provider has to analyze the content of the image and accomplish additional operations. In this paper, we propose a new robust vacate room after encryption schema based on polynomial arithmetic, which achieves a high embedding capacity with the perfect recovery of the original image. An efficient symmetric encryption method is applied to protect the privacy of the original image. One polynomial is generated by the encryption key and a group of the encrypted pixel, and the secret data is mapped into another polynomial. Through the arithmetic of these two polynomials, we can extract secret data and recover origin image, separately. Experimental results demonstrate that our solution has a stable and good performance on various images (include rough texture image). Compared with some typical methods, the proposed method can get better decrypted image quality with a large embedding capacity.
\end{abstract}


\end{frontmatter}

%\linenumbers

\section{Introduction}

Reversible data hiding (RDH) is a technology that enables the exact recovery of the original image when extracting embedded information. Due to the feature of lossless data hiding, it has gradually become a very active research area in the field of data hiding. The feature is desirable when highly sensitive data is embedding into an image, e.g., in military, medical, and legal imaging applications. The first RDH technique \cite{1999Method} was proposed by Barton in 2000 to address this concern. To date, various methods have been presented to achieve data hiding in images, in which difference expansion \cite{1227616} and histogram shift \cite{ni2006reversible} have far-reaching influence. Difference expansion is proposed by Tian in \cite{1227616}, the secret message is embedded by expanding the difference between adjacent pixels. In the histogram shifting method \cite{ni2006reversible}, the histogram of the original image is first generated, then it embeds messages by slightly modifying the pixel grayscale values of zero and the minimum point of the histogram. Subsequently, some schemes \cite{TSAI2013919,Local6746082,anew5648443,Efficient7122319} are introduced to improve the payload and reconstructed image quality based on them. Furthermore, to get better performance, methods such as lossless compression \cite{1608150Lossless9} and prediction error \cite{Wu_2014_22,1421361Prediction10} have been presented later.

As a great growth of cloud services and people's increasing attention for privacy protection, reversible data hiding in encrypted image (RDHEI) has drawn much interest from the research community. It can not only extract the secret message and recover the original image without error but also preserves the privacy of the original image. This concept was first proposed by Zhang \cite{5713234Zhang11} in 2011, there are three roles in the scenario, the image provider uses the encryption key to encrypt the image to protect the privacy of original images; a data hider embeds data by using the data hiding key without knowing the original image content; when having both above keys, the receiver is able to obtain the secret message and origin image correctly. It is helpful in some situations involving data privacy. For instance, in the cloud computing scenario, the user uploads an encrypted image, and an inferior assistant hopes to embed some message without access to the origin content.

Shi et al. \cite{7479451Reversible12} classify the existing RDHEI methods into two categories: vacating room before encryption (VRBE) and vacating room after encryption (VRAE). In VRBE methods \cite{6470679Ma13,Zhang_2014_14,7098386Cao15}, an extra preprocessing is performed before encryption to vacate space for data embedding. Ma et al. \cite{6470679Ma13} apply a traditional RDH method of histogram shifting to reserve room before encryption. The LSB of certain pixels is embedded into other pixels, which produces a pre-processed image. By substituting the LSB values of these vacated pixels in the encrypted image, the secret message can be embedded. The method can provide a larger payload (0.5 bpp) than previous methods with PSNR = 40 dB. In \cite{Zhang_2014_14}, based on the prediction technique, some pixels are estimated through the rest pixels before encryption. The secret message can be embedded by modifying the prediction errors instead of embedding data directly in encrypted images. By exploiting the correlation between neighbor pixels in nature images, Cao et al. \cite{7098386Cao15} use a sparse coding technique to free up a large space. The leading residual errors produced by sparse coding are encoded and then embedded into the original image. More secret message can be embedded into the vacated room in the encrypted image. In \cite{2020DoubleLi}, a new schema is proposed based on a double linear regression prediction model, which further improves the prediction accuracy and provides a large embedding capacity. Referring to the difference between current pixel and its predicted value, the secret message can be embeded. And they construct a prediction error map to record error positions for lossless recovering the original image. 

As VRBE is able to takes full use of spatial correlation between adjacent pixels, several high-cacity methods based on VRBE have been proposed \cite{2018An,2018High,2020Reversible} by compressing or using prediction error recently. For a pixel, Puteaux et al. \cite{2018An} substitute the MSB (most signiﬁcant bit) values to embed secret data. Then current pixel is predict by the decrypted previous ones. There may be some error in the recovered pixels, thus an error map is embeded to correctly reconstruct origin pixels. In \cite{2018High}, Chen et al. divide the image into blocks firstly. Duo to the spatial correlation of images, the MSB planes of a block have a large amount of consecutive 0s or 1s. Thus they design an algorithm called extend the run-length coding to compress MSB planes before encryption. In this way, the secret message can be embeded into the compressed planes. Yin et al. \cite{2020Reversible} use the common bits between a pixel and its predict value to embed message, and pixel labeling is applied to recover the orgin pixel from the predict value. A label map is recorded by using Huffman coding. Then the Huffman coding rule and the label map are sent as the auxiliary information. Finally, a high capacity can be achieved by by multi-MSB substitution.

For VRAE methods, the image provider directly encrypts the origin image, then the secret message are embedded into the encrypted image. The first VRAE method \cite{5713234Zhang11} is proposed by Zhang, the original image is firstly encrypted by a stream cipher, then the data hider divides the encrypted image into blocks and embeds secret bits by modifying a small proportion of encrypted data. The receiver can extract secret messages and recover the origin image perfectly by using the spatial correlation in natural images. Zhang \cite{6081934Zhang16} proposes a novel separable method, the exclusive-or operation is applied to encrypt the origin image, a sparse space to accommodate the secret data is created by compressing the least significant bits of the encrypted image. Qian and Zhang \cite{7076645Qian17} propose a novel method using distributed source coding. The image provider encrypts the image by using a stream cipher, then some bits selected from the encrypted image are compressed to make space for secret bits.

In existing RDHEI methods, the VRBE methods can provide a really high payload. However, an extra pre-processing operation is required before encryption for image provider. It may be unpractical, because the content owner needs to care about the the content of the origin image. And for most traditional reversible data hiding methods, data redundancy in natural 
images is applied to hide secret messages. Thus, it is difficult to embed messages into the encrypted image for VRAE method, since the correlation between a pixel and its adjacent neighbors is disappeared after encrypted. As expected, previous VRAE methods get limited embedding capacity or fail to recover the origin image. On the other hand, the performance of many exist methods is relate to the content of the image, they show a poor generalization performance on complex images (like Baboon image). In general, it is a challenge for RDHEI to achieve a high embedding capacity while recover the original image perfectly and have a stable performance on varies images.

Facing the challenge, we propose a new robust VRAE schema based on polynomial arithmetic. Our work is dedicated to finding a good solution not only in the recovery of the original image, but also in the embedding capacity. For the encryption of images, we design an efficient symmetric encryption method with support for correct data extraction and perfect image recovery in the RDHEI. The cipher of pixels in an image is generated via a univariate polynomial math expression $f(x)$, in which the coefficients of $f(x)$ are associated with both pixels and the encryption key, and the degree of $f(x)$ is the number of the pixels. The secret data is also mapped into another polynomial math expression $g(x)$. These expressions support polynomial addition, subtraction, and modulo operations to achieve data extraction and image recovery.

The rest of this paper is organized as follows. The proposed scheme is elaborated in Section 2. Abundant experimental results with the comparison and analysis are presented in Section 3. Finally, the conclusion is drawn in Section 4.

\section{Proposed Scheme}

\begin{figure}[h] %figure环境，h默认参数是可以浮动，不是固定在当前位置。如果要不浮动，你就可以使用大写float宏包的H参数，固定图片在当前位置，禁止浮动。
	\centering %使图片居中显示
	\includegraphics[width=1\textwidth]{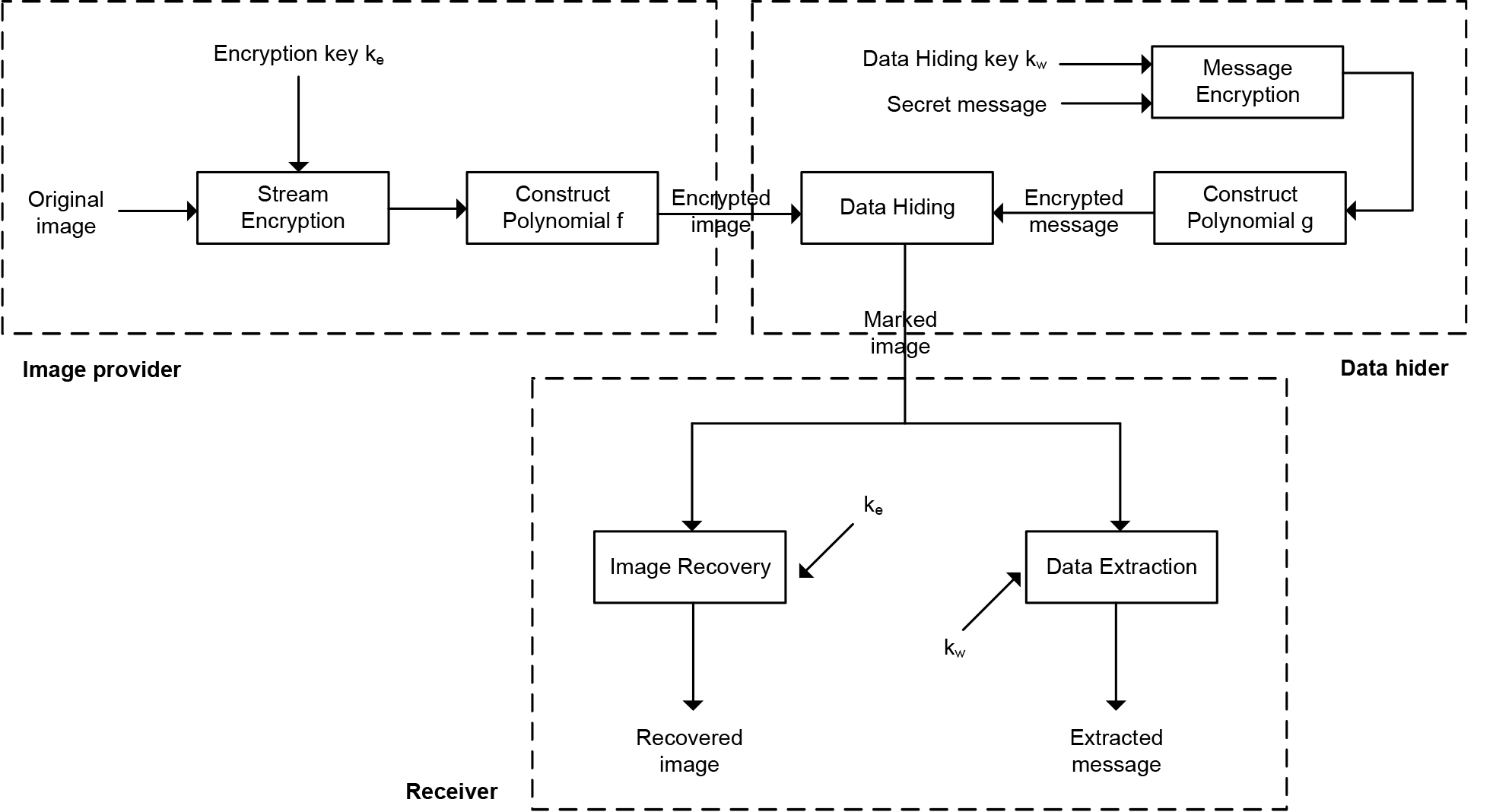} %中括号中的参数是设置图片充满文档的大小，你也可以使用小数来缩小图片的尺寸。
	\caption{Overview of the proposed schema} %caption是用来给图片加上图题的
	\label{fig:overview} %这是添加标签，方便在文章中引用图片。
\end{figure}

In the section, based on polynomial arithmetic, a novel solution is proposed to hide secret bits in pixels of origin image with low computational complexity and a high payload (the maximal payload could be 1.333 bpp for Lena). The overview of this process is shown in Fig. \ref{fig:overview}. 
%figure环境
There are three entities: an image provider, a data hider and a receiver. Stream encryption is applied to encrypted origin image firstly. Then the image is divided into blocks with the size of $2 \times 1$. For each group, the two pixels are converted with the key $k_e$ by the provider as a polynomial $f(x)$. Similarly, the hider using its key $k_w$  can encode two secret bits as a polynomial $g(x)$ with the same degree. Through the transaction of $f(x)’s$ coefficients, the encryption pixels are sent to the hider. Upon receiving the encrypted pixels, the hider computes $H(x)= f(x)+g(x)$. Note that the encrypted pixels hiding secret bits are coefficients of $H(x)$. Based on coefficients of $H(x)$ and the keys $k_e$ and $k_w$, the receiver can extract the secret bits and recover pixels of the original image perfectly.

\subsection{Image Encryption}
Assume that $k_e$ is an encryption key and $k_w$ is a data hiding key, where the former is shared between the provider and the receiver, the latter is shared between the hider and the receiver. And a public identity $ID$ is shared with everyone. 

Without loss of generality, assume the original image is of size $m \times n$. In this phase, a pseudo-random matrix C sized $m \times n$ is first generated by using $k_e$. For current pixel $x_{i,j}$ and its corresponding $c_{i,j}$, we convert them into the 8-bit binary sequence according to Eq. (\ref{toBits}), denoted as $x_{i,j}^k$ and $c_{i,j}^k$.
\begin{equation} \label{toBits}
	x_{i,j}^k=\lfloor \frac{x_{i,j}^k \bmod 2^{9-k}}{2^{8-k}} \rfloor\ , \quad k=1,2,\dots,8
\end{equation}
Then, current pixel is encrypted by following encryption operation.
\begin{equation}
	I_{i,j}^k=x_{i,j}^k \oplus c_{i,j}^k, \quad k=1,2,\dots,8
\end{equation}
Where $0 \le i < M$ , $0 \le j < N$, $I_{i,j}^k$ is the 8-bit binary sequence after encryption, and $\oplus$ denotes the exclusive-or(XOR) operation. Eventually, the encryped pixel can be obtained by Eq. (\ref{toDec}).
\begin{equation} \label{toDec}
	I_{i,j}=\sum_{k=1}^8 I_{i,j}^k \times 2^{8-k}, \quad k=1,2,\dots,8
\end{equation}

Next, divide the encryped image $I$ into blocks with the size of $2 \times 1$. Given a data list $(p_2, p_1)$ of two elements, where pixels $p_i (i=1,2)$ are selected from the above blocks. With a public identity $ID$, a polynomial $f(x)$ is constructed as follows
\begin{align}
f(x)&=p_2 (x+ID )^2+p_1 (x+ID)(\bmod \ 256) \nonumber \\
	&\triangleq a_2 x^2+a_1 x+a_0 (\bmod \ 256),
\end{align}
and the coefficient of function $f$ will be sent as the encrypted data. That is,
\begin{equation}
	E_{ID} (p_2,p_1 )=(a_2,a_1,a_0 )\triangleq \vec{a}.
\end{equation}
In this way, we get the final encrypted image.

\paragraph{Side information} Since the public identity is shared with everyone. It's more convenient to embed it into the encrypted image directly rather than network transmission. Because $ ID \in [2,15] $ (explained in section \ref{recover}), a 4-bit binary sequence $l$ is sufficient to represent it. For blind extraction and image recovery, we send $l$ as the side information. For all pixels, some of them are used to embed the side information, and the rest part to embed secret message. The strategy of sending side information is shown as
\begin{enumerate}[(1)]
	\item Records the origin LSB value of the first 4 pixels in encrypted image, denote as $\beta$.
	\item Replace the LSB of the first 4 pixels in encrypted image to save public identity bitstream $l$.
	\item Skip 4 pixels, $\beta$ will be embedded into the rest pixels by using our data embedding procedure.
\end{enumerate}

\subsection{Data Embedding}
In data embedding phase, the first 8 pixels are used to embed side information, then the boundary between first part and rest part can be obtained. The secret message will be embedded into second part of encrypted image without knowing the encryption key $k_e$. When having the encrypted image $\vec{a}$, the data hider first uses the $k_w$ to encrypt secret message $(m_2,m_1)$ :
\begin{equation}
	s_i=m_i \oplus k_i \ (i=1,2),
\end{equation}
where $k_i$ are obtained by $k_w$ using a standard stream cipher. Then they embed the encrypted secret bits $(s_2,s_1)$ into $\vec{a}$. Specifically, a polynomial $g(x)$ is generated as follows,
\begin{equation}
	g(x)=s_2 x^2+s_1 (\bmod \ 256).
\end{equation}
Then, the encrypted image embedded above secret bits is represented as
\begin{align}
	H(x)&=f(x)+g(x) \nonumber \\
		&=(a_2+s_2)x^2+a_1x+a_0+s_1(\bmod \ 256) \nonumber \\
		&\triangleq c_t x^2+c_1 x+c_0(\bmod \ 256).
\end{align}
Note that  $\vec{a}  = (a_2,a_1,a_0 )$. That is,
\begin{equation}
	E_{k_w} (s_2,s_1,\vec{a})=(c_2,c_1,c_0 )\triangleq \vec{c}.
\end{equation}

\subsection{Data extraction and image recovery}\label{recover}
Upon receiving the encrypted image embedded secret bits, the receiver extracts the secret bits and then recovers the original image. Specifically, the receiver works as follows,
\begin{enumerate}[(1)]
	\item Getting three pixels $(c_2,c_1,c_0)$ from the encrypted image embedded secret bits, she/he constructs a polynomial
	$$H(x)= c_2 x^2+c_1 x+c_0(\bmod \ 256).$$
	\item Collect LSB of first 4 pixels in the marked image, and  the binary sequence is converted to decimal to get public identity $ID$.
	\item Compute the value $H(-ID)$. Note that $f(-ID)=0$ and $H(-ID)=g(-ID)$.
	\item For $g(-ID)=s_2 \times ID^2 + s_1 (\bmod \ 256)$, the secret bits are extracted as follows
	\begin{align*}
		s_1&=[H(-ID)](\bmod \ ID)\\
		s_2&=\frac{H(-ID)-s_1}{ID^2}
	\end{align*}
	\item Based on the extracted secret bits, the polynomial g(x) is obtained as
	$$g(x)=s_2 x^2+s_1(\bmod \ 256).$$
	\item For $i=0,1,2$, let $a_i= (c_i-s_i)\bmod\ 256$. Then, $f(x)$ can be reconstructed as
	$$f(x)=a_2 x^2+a_1 x+a_0 (\bmod\ 256).$$
	\item By using $(a_2,a_1,a_0 )$, the receiver can recover encrypted pixels $p_2$ and $p_1$ as follows,
	\begin{align*}
		p_2&=a_2 \\
		p_1&=a_1-p_2*C_2^1*ID (\bmod\ 256)
	\end{align*}
	\item From the extracted secret bits, $\beta$ and real secret message could be decomposed.
	\item Decrypting secret message by using $k_w$.
	\item With the origin LSB $\beta$, the first 4 encrypted pixels can be recovered by replacing their LSB value.
	\item When having $k_e$, the original pixels can be recovered free of error as
	$$x_{i,j}=I_{i,j} \oplus c_{i,j}$$

\end{enumerate}
As previously mentioned, the encrypted image and encrypted secret data could be split by using $ID$, thus we can reconstrut the origin image and extract secret message separately. With different secret keys the receiver owns, there are three possible cases:
\begin{enumerate}
	\item The receiver only has the encryption key $k_e$: After obtaining the split encrypted image, a same pseudo-random matrix C sized $m \times n$ is first generated by using the encryption key, then each origin pixel can be recovered according to step (11).
	\item The receiver only has the data hiding key $k_w$: With the data hiding key $k_w$, the receiver decrypts the split encryped message to extract the origin secret message correctly.
	\item The receiver has both the encryption key $k_e$ and the data hiding key $k_w$: We can recover origin image perfectly and extract secret message correctly.
	
\end{enumerate}

\begin{figure}[h]
	\centering 
	\includegraphics[width=0.6\textwidth]{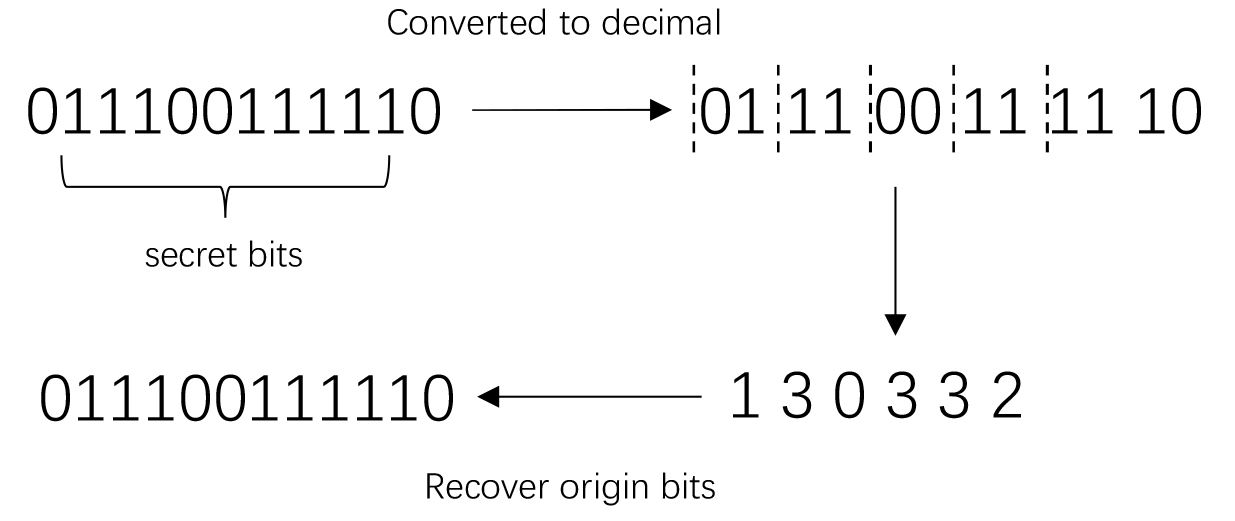}
	\caption{Secret data processing} 
	\label{fig:secret} 
\end{figure}

\paragraph{Notice} In step (4), for correctly extracting secret message, the value of $ID$ is limited: 
\begin{align*}
	\begin{cases}
		ID>s_1 \\
		s_2 \times ID^2<256 
	\end{cases}
\end{align*}
From the above inequations, it concluded that two bits can be embeded into each coefficient of $f$ at most. The details cases are as follows:
\begin{enumerate}
	\item when embedding 1 bits, $s_i \in [0,1] \ (i=1,2)$ and $ID \in [2,15]$.
	\item when embedding 2 bits, $s_i \in [0,3] \ (i=1,2)$ and $ID \in [4,9]$. In this case, secret message need to be extra processed during different phase. Before data hiding, the data hider have to divide the secret bits into a group of 2 bits, and then convert them to decimal similar to Eq. (\ref{toDec}). Furthermore, in data extraction phase, the receiver needs to convert the extracted decimal secret data to origin secret bits, as indicated in Fig. \ref{fig:secret}.

\end{enumerate}

\begin{figure}[h]
	\centering 
	\includegraphics[width=1\textwidth]{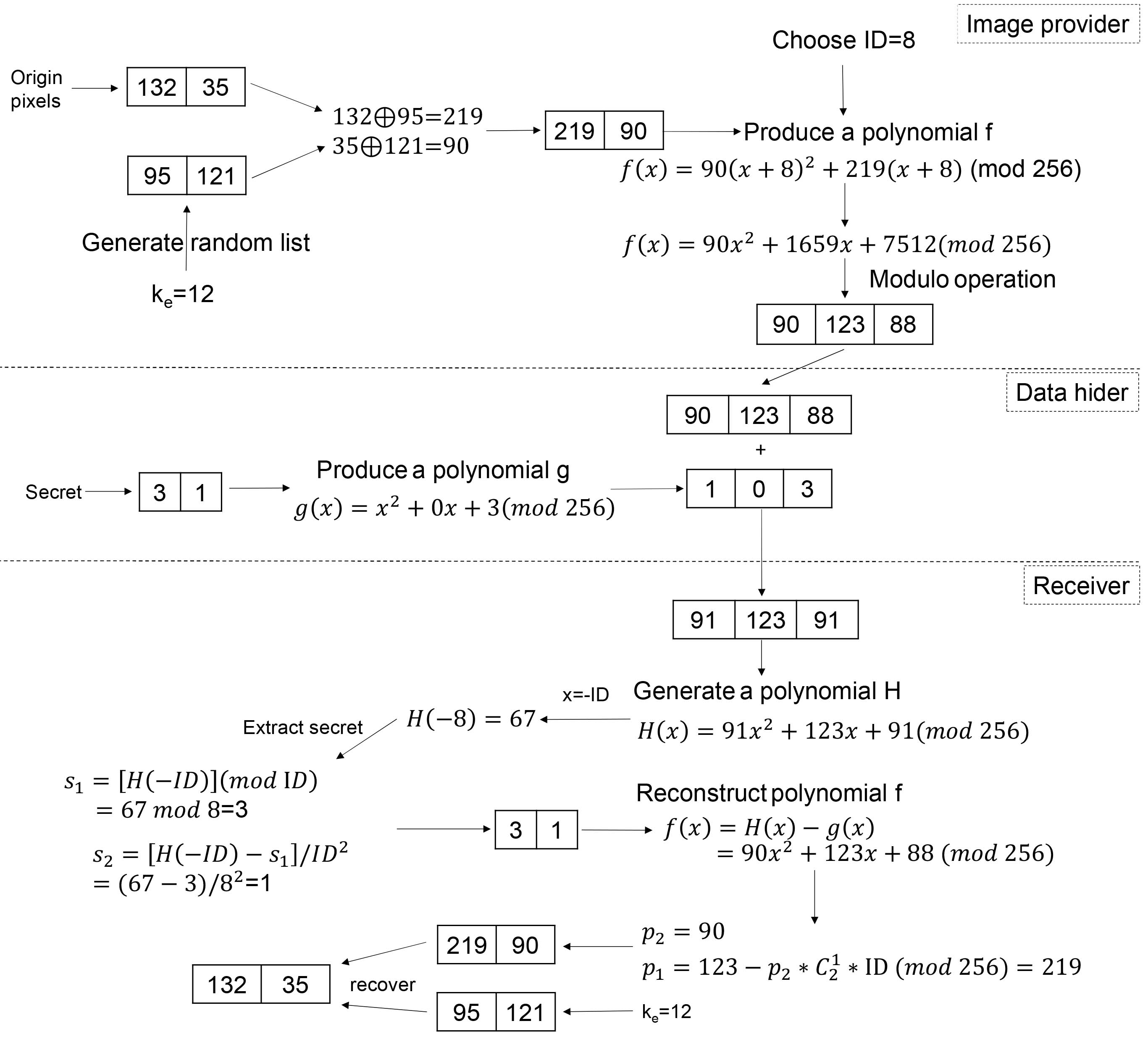}
	\caption{Example of the proposed schema} 
	\label{fig:example} 
\end{figure}

\paragraph{Example 1} Fig. \ref{fig:example} shows the sketch of this example.
%figure环境
 For an 8-bit image, given two pixels (132,35) and $k_e=12$, the image provider firstly generates a pseudo-random list (95,121) by $k_e$. The original pixels are encrypted as (219,90) by exclusive-or operations. We randomly choose 8 as $ID$. Then the provider produces a polynomial $f(x)=90x^2+1659x+7512(\bmod\ 256)$. The final encrypted pixels can be obtained by modular operation, set the output (90,123,88).

In the embedding data phase, with the secret message $s_1=3$, $s_2=1$ that are encrypted by $k_w$, the hider also generates a polynomial $g(x)=x^2+3(\bmod\ 256)$, and gets the output (1,0,3) by the modular operation. Finally, due to the additive modular arithmetic, the encrypted pixels containing a secret message (91,123,91) are obtained by adding (90,123,88) and (1,0,3).

When receiving the encrypted pixels, the receiver extracts $ID=8$ and generates a polynomial $H$. Let $x=-ID$, and get $H(-8)=67$. Then receiver uses $ID$ to extract the hidden message $s_1=3$ and $s_2=1$. With the extracted message, the receiver reconstructs the polynomial $f$. From the coefficient of $f$, the origin pixels (132,35) could be recovered with $k_e=12$.

\section{Experimental results}
\begin{figure}[]
	\centering 
	\includegraphics[width=0.75\textwidth]{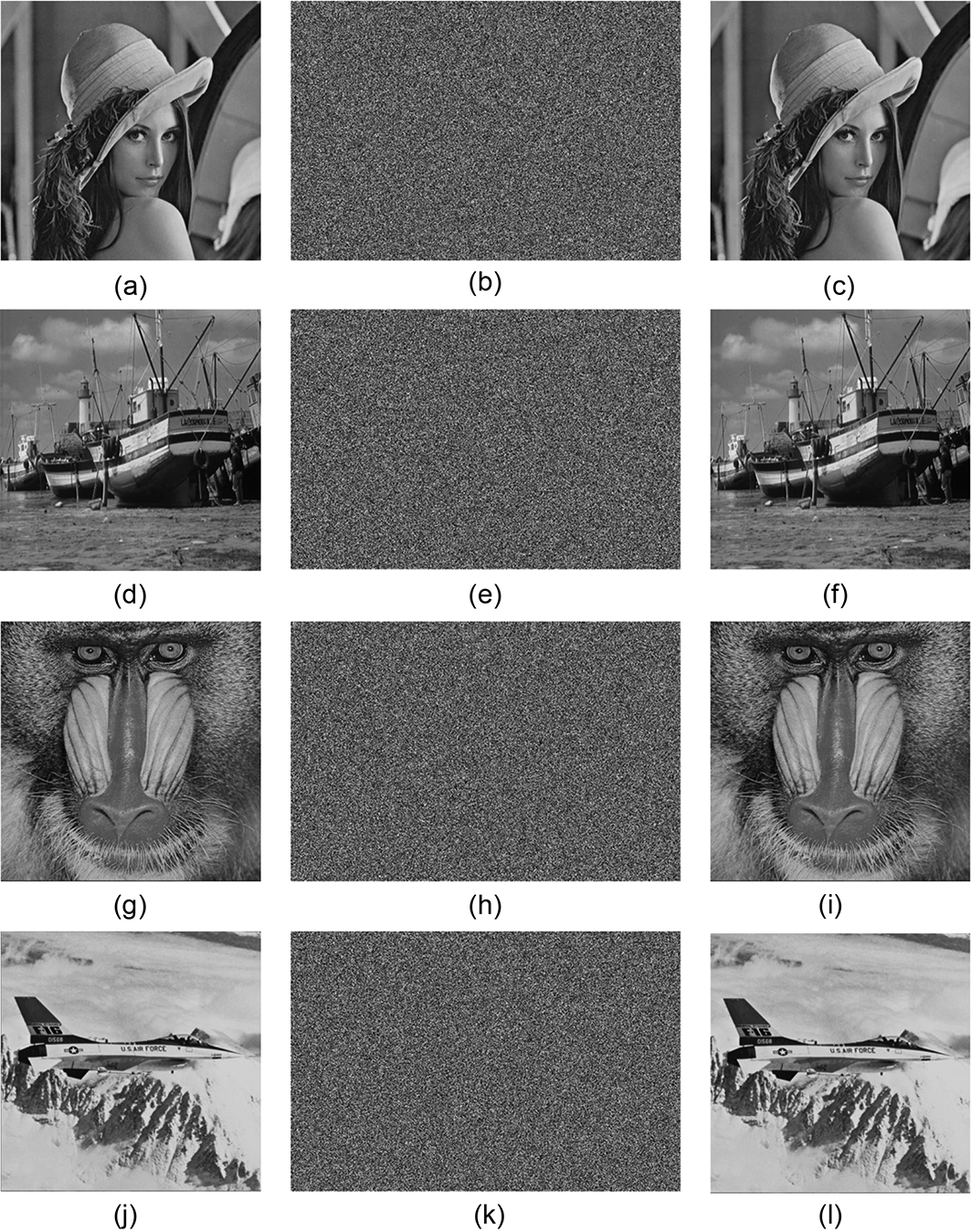}
	\caption{Test images with 512 × 512 pixels. Left: Original-image. Middle: Encrypted with secret message, Right: After Decryption.} 
	\label{fig:images} 
\end{figure}

In this section, based on the embedding capacity (payload) and reconstructed image quality, we evaluate the performance of the proposed method. The method is tested on the uncompressed 512×512 gray-scale images chosen from the USC-SIPI image database\cite{USC-SIPI-20}. To validate the proposed scheme, experiments are done with a variety of standard test gray-scale images. Some are presented in Fig. \ref{fig:images}  namely Lena, Boat, Baboon, and Airplane. The comparisons between the existing methods and ours are shown in Fig. \ref{fig:lena} and \ref{fig:Baboon}, according to the embedding rate (EC) and the peak-signal-to-noise ratio (PSNR) of images. Here, EC is adopted to evaluate the embedding capacity, and it calculated as
\begin{equation}\label{con:ec}
	EC=\frac{Total\ embedded\ bits}{Total\ number\ of\ pixels\ in\ a\ image}.
\end{equation}
Again, the PSNR is applied to evaluate the reconstructed image quality in comparison to the original image, which is computed as follows.
\begin{align}\label{con:psnr}
	MSE=\frac{1}{n}\sum_{i=1}^n(p_i-p_i')^2 \nonumber \\ 
	PSNR=10\log_{10}{\frac{255^2}{MSE}}.
\end{align}
where $n$ is the total number of pixels, $p_i$ and $p_i'$ denote the pixel values of origin and recovered images, respectively. Finally, we would compare our schema with some related works, and discuss their efficiency.

%figure环境

\subsection{Performance Analysis}
The proposed method can perfectly recover the original images from the corresponding encrypted images as shown in Fig. \ref{fig:images}. In our experiments, 200000-bits is embedded into four randomly selected images. The original image is shown at the left of the figure. After encryption and data embedding, an encrypted image with the secret message shown in the middle of the figure is generated with the 8 encrypted bits. The rest is the decrypted image. The comparison shows the pixel values of the recovered image is the same as that of the original image (PSNR$\to+\infty$), which indicates the images could be perfectly recovered. The corresponding experimental results are shown in Table \ref{tab:0}.

\begin{table}[]
	\caption{EXPERIMENTAL RESULTS OF FOUR SELECTED IMAGES}
	\centering
	\label{tab:0}   
	\begin{tabular}{llllll}
		\hline
		& Total capacity(bits) & EC(bpp) & PSNR & SSIM & Extra bits    \\
		\hline
		Lena   & \makecell[c]{524280}     & 1.333  & $+\infty$   & \makecell[c]{1}  & \makecell[c]{4} \\
		Boat   & \makecell[c]{524280}    & 1.333  & $+\infty$   & \makecell[c]{1} & \makecell[c]{4}      \\
		Baboon  & \makecell[c]{524280}    & 1.333  & $+\infty$   & \makecell[c]{1} & \makecell[c]{4}     \\
		Airplane & \makecell[c]{524280}   & 1.333  & $+\infty$   & \makecell[c]{1}  & \makecell[c]{4}     \\
		\hline
	\end{tabular}
\end{table}

Moreover, we make a performance testing by a random selection of 1000 grayscale images with size 512x512 on BOWS-2 \cite{BOWS-2_21} and BOSSbase \cite{Bossbase} datasets. The test results for these images are shown in Table \ref{tab:1}. For embedding capacity, the best case is 1.333 bpp, and 1.333 bpp in the worst case. It is clear that the worst EC equal to the best EC. And the average EC can be also reach 1.333 bpp, which indicate that the payload of all image is 1.333 bpp. Certainly, each original image could be reconstructed free of error (satisfy PSNR$\to+\infty$ and SSIM=1). From the above analysis, it indicates that the proposed method has stable and good performance.
\begin{table}[]
	\caption{PERFORMANCE MEASUREMENTS OF 1000 IMAGES ON DATASETS}
	\centering
	\label{tab:1}   
	\begin{tabular}{lllll}
		\hline
			Dataset & Indicator & Best case & Worst case & Average      \\
		\hline
		
		\multirow{3}{*}{BOWS-2}
		& EC (bpp)  & 1.333      & 1.333      & 1.333        \\
		& PSNR (dB) & $+\infty$  & $+\infty$  & $+\infty$     \\
		& SSIM      & 1          &  1         &1     \\
		
		\hline
		
			\multirow{3}{*}{BOSSbase}
		& EC (bpp)  & 1.333      & 1.333      & 1.333        \\
		& PSNR (dB) & $+\infty$  & $+\infty$  & $+\infty$     \\
		& SSIM      & 1          &  1         &1     \\
		
		\hline
	\end{tabular}
\end{table}

\begin{figure}[]
	\centering 
	\includegraphics[width=0.75\textwidth]{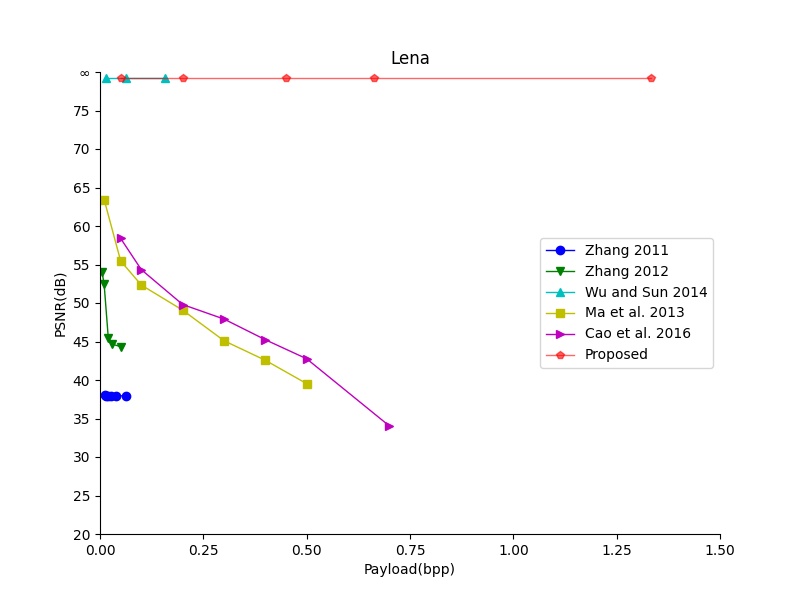}
	\caption{Performance comparison on Lena} 
	\label{fig:lena} 
\end{figure}

\begin{figure}[]
	\centering 
	\includegraphics[width=0.75\textwidth]{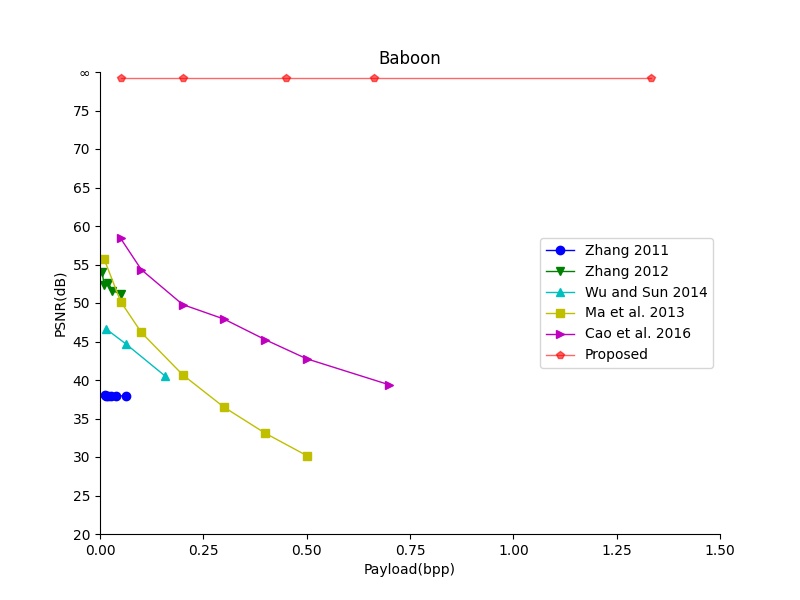}
	\caption{Performance comparison on Baboon} 
	\label{fig:Baboon} 
\end{figure}

We now compare the performance of our approach with some classic design \cite{5713234Zhang11,6470679Ma13,7098386Cao15,2020DoubleLi,6081934Zhang16}. Our experiments are based upon the two typical images of Lena and Baboon. Then, their embedding capacity and PSNR are estimated through (\ref{con:ec}) and (\ref{con:psnr}). The results are summarized in Fig. \ref{fig:lena} and Fig. \ref{fig:Baboon}, respectively. As expected, across all six designs and in terms of two criteria considered here, the proposed approach exhibits the best PSNR. When the embedding capacity increases, our approach has a stable performance on PSNR compare to others. Thus, when the receiver having the encryption key and without the data hiding key, our approach is the only schema that can perfectly recover both Lena and Baboon image (PSNR$\to+\infty$). As for EC, our method can reach 1.333 bpp for both Lean and Baboon. It is clear that our method can obtain a higher payload than other related methods, except for Cao et al \cite{7098386Cao15} and Li et al \cite{2020DoubleLi}. These show that our approach provides a good trade-off between the embedding capacity and PSNR.
%figure环境
%figure环境

\begin{figure}[]
	\centering 
	\includegraphics[width=0.75\textwidth]{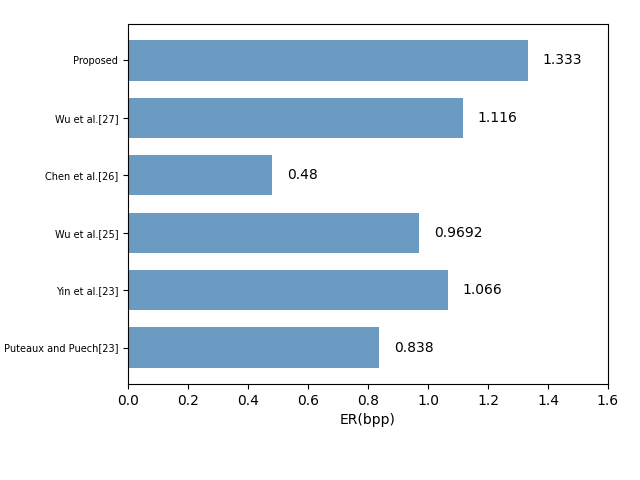}
	\caption{ER comparison on Baboon with high-capacity VRBE methods} 
	\label{fig:cmp_baboon}
\end{figure}

To prove the robustness of the proposed scheme, we compare our approach with some state-of-the-art high-capacity VRBE methods \cite{2018An,2020Reversible,2019An,2018High,2021Secure} on complex image Baboon. For all these methods can recover origin image losslessly (PSNR$\to+\infty$), we only need to make comparasion on embedding rate(ER) here. The result is  illustrated in the Fig. \ref{fig:cmp_baboon}. The ER of the proposed method can reach 1.333 bpp. It can be seen that these high-capacity methods get lower ER than the proposed schema. In these high-capacity methods, a pre-processing is required to vacate room for data embedding, thus the ER is closely relate to the content of origin image. It can lead to poor performance of these methods on complex images. In contrast, the proposed method directly encrypt origin image and has a very stable performance on all images. This validates that the proposed method is robust.

%figure环境

\subsection{Security Analysis}
In the proposed method, it is able to protect the security of the origin image and secret message by using effective encryption algorithm. To prove the security of the proposed schema, we will mainly focus on the applied encryption method and statistical analysis including histogram and correlation analysis in this section. In the image encryption phase, the origin image is directly encrypted with a stream cipher. For a image of size $m \times n$, a random matrix C is first generated. Each element of C is converted into an eight-bit binary number, thus the guessing probability of the secret key
is $ \frac {1}{2^{8 \times m \times n}}$, which is too low to predict the correct secret key. As for secret message, it can be encrypted with a stream cipher like image encryption or any secure symmetric encryption algorithms. Hence, it is impossible to get the origin image or secret message without corresponding secret key.

\begin{figure}[]
	\centering 
	\includegraphics[width=1\textwidth]{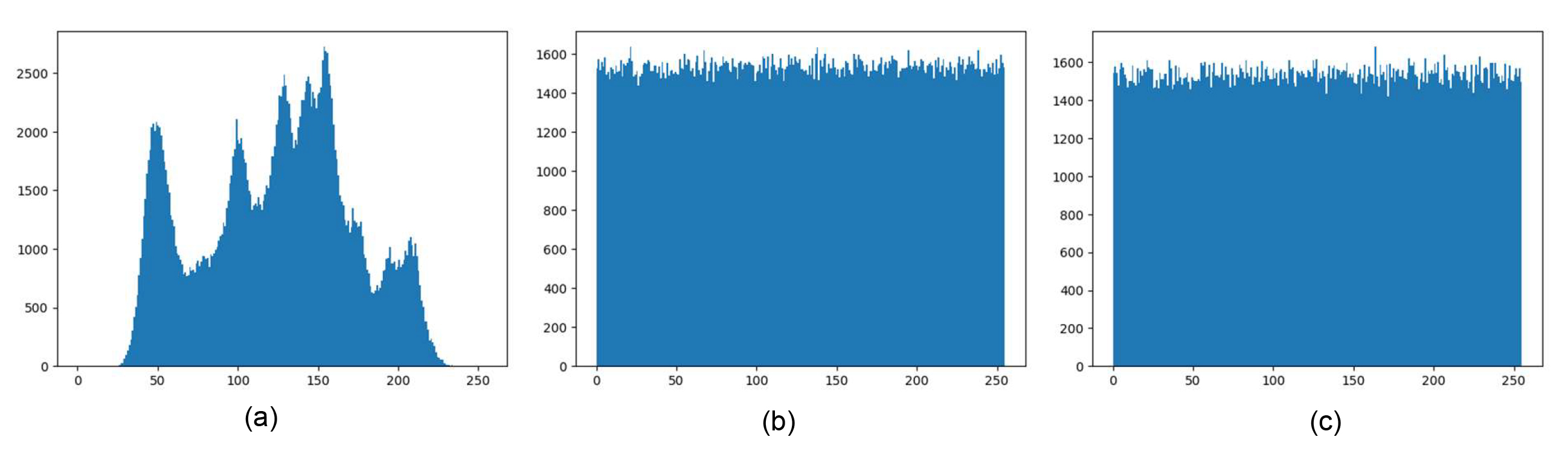}
	\caption{Histograms of Lena image on three stages. (a) Origin image (b) Encrypted image (c) Marked encrypted image} 
	\label{fig:hf}
\end{figure}

Take Lena image to illustrate security analysis. Fig. \ref{fig:images} (a) is the origin image and Fig. \ref{fig:images} (b) is the marked encrypted image. It is clear that the encrypted image is noise-like and no useful information can be obtained from the image. Furthermore, we analyze the histogram, correlation coefﬁcient and entropy of Lena to further validate the security of the proposed method. Fig. \ref{fig:hf} shows the histograms of the origin image, the encrypted image and the marked image. From these histograms, it can be found that the histogram of the origin image has meaningful information, while the pixel distribution of the encrypted image and the marked image are both uniform compare with the original image . It is difficult to obtain the origin image from the useless statistical feature. This validates that the proposed method is guaranteed a high security level.
%figure环境

\subsection{Data Expansion}
Besides, to measure the efficiency of the proposed schema, we make a comparison with other schemas on data expansion. The result is given in Table \ref{tab:2}. Data expansion indicates the encrypted image is bigger than the original image. We denote the number of origin pixels as $N$. There are some methods \cite{6081934Zhang16,6151038Hong23} using stream cipher encryption to encrypt images, which adopt exclusive-or operation to encrypt images. It obviously does not cause data expansion. For those methods \cite{8013725Xiang24,Shiu_2015_25} using Paillier homomorphic encryption, a 512-bits or 1024-bits secret key is customarily used to encrypt 8-bits pixels. After encrypted by the Paillier encryption, the origin 8-bits pixel is at least expanded to 1024-bits, thus the encrypted pixel is further expanded. The rest schemas \cite{8705382Chen18,Wu_2018_19} apply secret sharing to encrypt images, method \cite{Wu_2018_19} would expand the original image to two or more times. In method \cite{8705382Chen18}, they improve secret sharing into multi-secret sharing to prevent data expansion. As for our approach based on polynomial arithmetic, two pixels would be encrypted into three pixels, so the encrypted image expands 1.5 times.
\begin{table}[]
	\caption{COMPARISONS OF DATA EXPANSION}
	\centering
	\label{tab:2}   
	\begin{tabular}{llllll}
		\hline
		\makecell[c]{Schema} & \makecell[c]{Encryption method} & \makecell[c]{Encryption space}   \\
		\hline
		\makecell[c]{[16,23]}  & \makecell[c]{Stream cipher}     & \makecell[c]{$N$}    \\
		\makecell[c]{[24,25]} & \makecell[c]{Paillier encryption}  & \makecell[c]{$\le 128N$}     \\
		\makecell[c]{[19]}      & \makecell[c]{Secret sharing}          & \makecell[c]{$\le 2N$}  \\
		\makecell[c]{[18]}      & \makecell[c]{Secret sharing}          & \makecell[c]{$N$}         \\
		\makecell[c]{Ours}      & \makecell[c]{Polynomial arithmetic}          & \makecell[c]{$1.5N$}         \\
		\hline
	\end{tabular}
\end{table}

\section{Conclusion}
In this work, a new method of reversible data hiding in encrypted images based on polynomial arithmetic is presented, which achieves a good balance between embedding capacity and reconstructed image quality. Since the correlation of adjacent pixels is disappear after encryption, thus various methods proposed is trying to preserve the feature to vacate room for embedding message. Specially, we propose a new method to avoid utilizing the correlation of adjacent pixels in nature images, thus our methods is robust and has good performance on all images. Through arithmetic of the two polynomials, a high capacity can be obtained while the reconstructed image is the same as the original image. However, data expansion may raise storage costs for encrypted images. In the future, we will try to solve the problem of data expansion and design more efficient polynomials to further improve the embedding capacity.

%\section*{References}

\bibliography{mybibfile}

{\renewcommand*\numberline[1]{Fig.\,#1:\space}
	\makeatletter
	\renewcommand*\l@figure[2]{\noindent#1\par}
	\makeatother
	
	\listoffigures}

\end{document}